\documentclass[10pt,letter,twoside]{rho}
\usepackage[english]{babel}

\setbool{rho-abstract}{true} %
\setbool{corres-info}{true} %

\title{Artificial collectives of specialists and generalists excel at different tasks}

\author[1]{John Meluso}
\author[2,3]{Laurent H\'ebert-Dufresne}
\author[4]{Christoph Riedl}
\author[1]{H. Oliver Gao}

\affil[1]{Cornell University}
\affil[2]{University of Vermont}
\affil[3]{Santa Fe Institute}
\affil[4]{Northeastern University}

\dates{This manuscript was compiled on \today}

\leadauthor{Meluso et al.}
\footinfo{Creative Commons CC BY 4.0}
\smalltitle{Artificial collectives of specialists and generalists excel at different tasks}
\theday{\today} %

\corres{John Meluso}
\email{jam627@cornell.edu}

\license{This document is licensed under Creative Commons CC BY 4.0.}

\begin{abstract}
    Collective artificial intelligence---where multiple agents work on shared tasks---holds potential to solve expansive problems in fields from medicine to collective governance.
    But while prescriptive engineering solutions abound, we lack descriptive scientific understanding of artificial collectives, and therefore principles for how to design resource efficient multi-agent systems.
    Through systematic experiments with optimizing agents, we characterize how agent interpretive abilities, rationality bounds, and task qualities interact to shape collective performance.
    Agents range from specialists, with narrow interpretive abilities, to generalists, with broad ones.
    Collectives of specialists correspond to sparse, centralized networks, while collectives of generalists correspond to dense, decentralized ones.
    We show that interpretive network properties have small performance effects on average (0.07 standard deviations of performance).
    However, for specific task qualities, these effects are 4.5 times larger (0.33 sd) and can reach much higher for certain task qualities (1.84 sd).
    This leads collectives of generalists to perform better on tasks that involve generating, choosing, and coordinating, while collectives of specialists with a few generalist mediators perform better on tasks that involve negotiating.
    Rationality bounds then moderate these relationships.
    At loose bounds, specialists outperform generalists through more effective sampling of high-dimensional decision spaces. 
    At tight bounds, generalists outperform specialists through better gradient estimation.
    A fundamental trade-off between performance and convergence speed emerges at moderate bounds.
    These findings suggest that effective multi-agent design could benefit from matching interpretive networks to both task demands and agents' computational limits, with likely implications for the efficiency and energy costs of multi-agent systems.
\end{abstract}

\keywords{collective intelligence, multi-agent systems, bounded rationality, network science, distributed problem-solving, artificial general intelligence}

\begin{document}
	
    \maketitle
    \thispagestyle{firststyle}
    \renewcommand{\dblfloatpagefraction}{0.7}

\section*{Introduction}

\begin{figure}
    \centering
    \includegraphics[width=\linewidth]{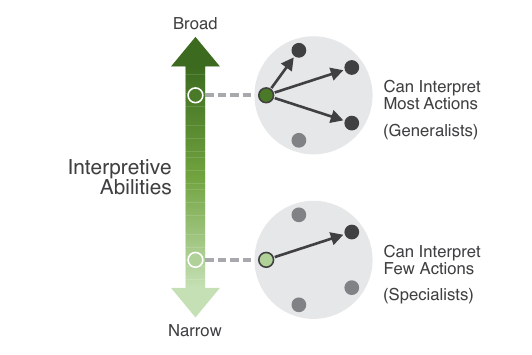}
    \caption{
        \textbf{Interpretive abilities correspond to network ties.}
        Agents can have many different interpretive abilities.
        Varying how many interpretive abilities an agent has implies a conceptual spectrum, from a narrow set of abilities (can interpret few actions, like specialists) to a broad set (can interpret varied actions, like generalists).
        With respect to a group, having greater interpretive abilities corresponds to more network ties while having fewer corresponds to fewer network ties.
    }
    \label{fig:interpretation}
\end{figure} 
Collective intelligence in systems of artificial agents is a growing scientific frontier with significant implications for problem-solving across domains \cite{Riedl2025AI, Sehwag2025Collective}.
These multi-agent systems demonstrate remarkable capabilities in strategy games \cite{Silver2018General}, scientific discovery \cite{Strieth-Kalthoff2024Delocalized}, and spatial navigation \cite{Rahwan2019Machine}.
Their capabilities hold potential to assist humans with challenging medical decisions \cite{Kim2024Demonstration}, engineering optimization \cite{Chen2024SoS, Nitti2025Collective}, collective governance \cite{Barfuss2025Collective, Tacchetti2025Deep}, and many other complex tasks.
At the same time, multi-agent systems are substantial contributors to growing datacenter resource consumption and carbon emissions \cite{Economist2024BigTech, IEA2024Electricity, Leppert2025What, IEA2025Energy, Aczel2026Environmental}.
Together, these pros and cons make it essential to answer: \textit{How should we design multi-agent systems to take advantage of their benefits while mitigating their drawbacks?}
Answering such questions requires not just prescriptive engineering of solutions but a descriptive science of artificial systems \cite{Simon1996Sciences, Rahwan2019Machine}.
This paper systematically investigates two understudied design qualities of artificial systems: agent interpretive abilities \cite{Gronauer2022Multiagent, Oroojlooy2023Review} and bounded rationality \cite{Simon1957Models}.

The first underexplored quality of these systems is the spectrum of agent interpretive abilities (Fig. \ref{fig:interpretation}) \cite{Gronauer2022Multiagent, Oroojlooy2023Review}.
Interpretive abilities enable agents to encode and decode messages \cite{Shannon1948Mathematical}.
In turn, this allows agents to form mental models of others because agents can productively understand others' actions and share information in interpretable ways \cite{Westby2023Collective, Kelley2025Personalized}.
However, agent interpretive abilities vary substantially.
Agents with narrow interpretive abilities can interpret few others' actions and are themselves interpretable by few.
In contrast, agents with broad interpretive abilities can interpret many others' actions and can share information to make themselves interpretable by many.
This range of interpretive abilities corresponds to networks among agents in which ties describe \textit{who can effectively interpret whom} \cite{Horling2004Survey}, just as social perceptiveness shapes ties in human groups \cite{Amelkin2018Dynamics, Centola2022Network}.
Such abilities also map onto the specialist-generalist spectrum of skills studied in human collective intelligence research \cite{Hong2004Groups, Goldstone2024Emergence}.
Specialists develop expertise in a narrow set of skills at the cost of interpretive breadth, while generalists develop a broad set of interpretive abilities but less depth in any domain.

\begin{figure*}
    \centering
    \includegraphics[width=7.5in]{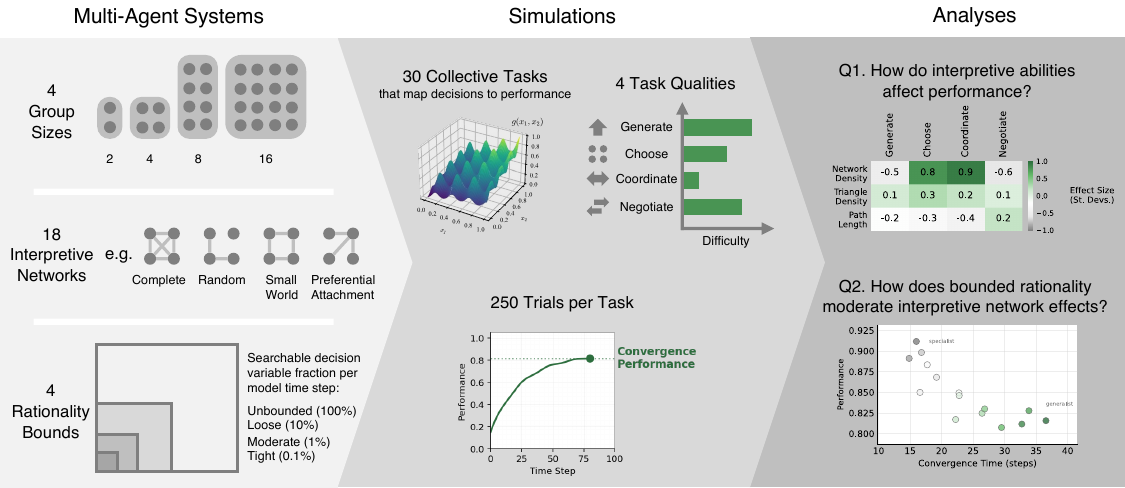}
    \caption{
        \textbf{Methodological overview.}
        We systematically varied multi-agent system design properties including group size (2, 4, 8, 16 agents), interpretive networks (18 topologies including complete, random, small-world, and preferential attachment graphs), and agent rationality bounds (searchable fraction of decision variable domain of $\pm0.1\%$, $1\%$, $10\%$, $100\%$ per time step).
        Each multi-agent system sought to maximize performance on one of 30 collective tasks (mathematical objective functions mapping states to performance).
        Task qualities varied significantly as measured by 4 difficulty measures (generating useful new solutions, choosing the best option, coordinating decisions, and negotiating with competing objectives).
        We measured convergence performance across 250 trials per task.
        Analysis addressed two research questions: (Q1) What interpretive network properties affect performance across task types? We used linear regressions to quantify network property and task quality effects.
        (Q2) How does bounded rationality moderate interpretive network effects? We compared performance differences across network densities for systems with varying rationality bounds.
        Equal computational resources (32 decision variables) were maintained across all group sizes.
    }
    \label{fig:methods}
\end{figure*} 
Google DeepMind's AlphaStar exemplifies this interpretive spectrum \cite{Vinyals2019Grandmaster}.
Using about 900 co-trained agents, the system achieved grandmaster-level performance in StarCraft II, a challenging strategy game.
Training patterns shaped each agent's interpretive abilities: specialized ``exploiter'' agents trained against limited strategies, learning to interpret only those approaches and thus developing narrow interpretive abilities.
Meanwhile, generalist ``main'' agents trained against diverse opponents, developing broad interpretive abilities that achieved grandmaster-level performance.
AlphaStar's success depended on collaboration among agents with different interpretive abilities---narrow exploiters and broad main agents.
Such differences in interpretive abilities correspond to networks among agents where ties represent who can effectively interpret whom.
Similar interpretive heterogeneity appears throughout multi-agent systems research \cite{Tan1993Multiagent, Stone2000Multiagent, Shoham2008Multiagent, Strieth-Kalthoff2024Delocalized}.

Network topologies influence group performance in human collective intelligence research, with a group's balance of interpretive abilities shaping effectiveness \cite{Bavelas1950Communication, Bavelas1965Experiments, Grant1996Toward, Bunderson2002Comparing, Rulke2000Distribution, Postrel2002Islands, Centola2022Network, Meluso2023Multidisciplinary}.
Network topologies also influence performance on tasks with different qualities such as generate, choose, negotiate, and coordinate tasks \cite{McGrath1984Groups, Lazer2007Network, Mason2012Collaborative, Barkoczi2016Social}.
Whether these dynamics of human groups transfer to artificial systems remains a largely open question in collective intelligence research, though.

Despite growing deployment of multi-agent systems, we lack scientific understanding of how interpretive networks between artificial agents influence collective performance across different tasks.
Current approaches to multi-agent systems use various broadcasting and targeted interaction topologies \cite{Gronauer2022Multiagent, Horling2004Survey} but seldom study the appropriateness of interpretive network topologies for specific tasks.
Additionally, current approaches rarely examine how network design should be informed by agents' inherent computational limits---what Herbert Simon called bounded rationality \cite{Simon1957Models}.

This leaves two critical knowledge gaps.
First, while the effects of network topology on group performance are well-established in human research, we lack descriptive evidence of how interpretive network topologies affect collective performance among \textit{artificial} agents in controlled settings.
Second, and more fundamentally, we have limited understanding of how interaction between network topologies and agents' computational limits affect collective performance.
These gaps pose fundamental scientific challenges because, just as insufficient knowledge of physics would limit effective bridge design, lacking a science of artificial collective intelligence limits our ability to design effective multi-agent systems.
Our research addresses two exploratory questions:
\begin{enumerate}
    \item What interpretive network properties most improve multi-agent group performance, both overall and for specific task qualities?
    \item How does bounded rationality moderate the relationship between interpretive network density and collective performance?
\end{enumerate}

To answer these questions, we designed an experiment that systematically varies key properties of multi-agent systems, measuring their performance across a battery of 30 varied mathematical representations of tasks (see Fig. \ref{fig:methods}).
Instead of using specific AI implementations (like particular large language models or other machine learning implementations), we model agents abstractly as optimizers that iteratively search constrained state spaces.
This abstraction captures how researchers and practitioners predominantly design agentic systems to maximize performance by searching state spaces within computational limits \cite{Russell2020Artificial, Silver2021Reward, Gershman2015Computational} (see Methods).
Tasks are represented by objective functions that map agents' collective states to performance across high-dimensional landscapes. The landscapes vary along four non-mutually exclusive qualities derived from organizational psychology \cite{McGrath1984Groups, Meluso2023Multidisciplinary, Hu2025Task}: generate, choose, coordinate, and negotiate, each capturing a distinct source of collective difficulty (see Methods).

Our results show that collective performance depends on appropriately matching interpretive network topology with both task qualities and rationality bounds.
Task qualities produce effects 4.5 times larger than network properties alone.
Interpretive network density (the proportion of agent pairs that can effectively understand one another), decentralization, and path length can have effects 20 times larger (nearly 2 standard deviations) but vary significantly by task.
We also find that bounded rationality has mixed effects, favoring groups of specialists in some cases (loosely bounded, +6.7\%) and generalists in others (tightly bounded, +7.3\%).
Between the extremes, moderately bounded searches pose a trade-off between speed and quality with conditions that will require practitioner testing and greater scientific investigation.
Together, these findings suggest that efficient multi-agent system designs must be specific to task qualities and agent computational limits, rather than universal.

\section*{Results}
\subsection*{Different tasks benefit from different interpretive network properties}
\begin{figure*}
    \centering
    \includegraphics[width=7.5in]{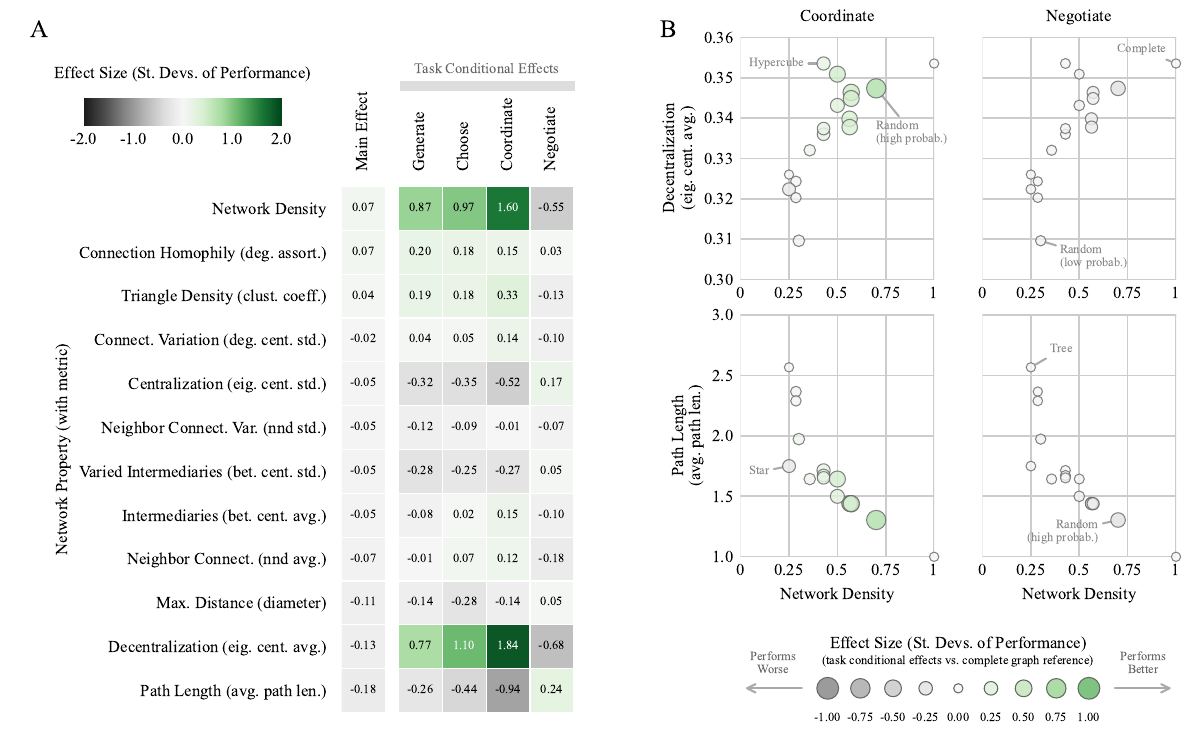}
    \caption{
        \textbf{Performance effects of interpretive network properties.}
        (A) A heatmap showing the performance of different network properties in terms of standard deviations above or below the mean of all simulation runs.
        Values show network property effects, both across tasks qualities (the main effects) and limited to specific task qualities (the task conditional effects).
        (B) The performance of specific network topologies corroborate the large effects seen for decentralized and dense networks with short paths between nodes (task conditional effects for teams of $n=8$).
        All values shown are statistically significant multivariate regression results ($p<0.05$).
        Higher graph density corresponds to greater average agent interpretive abilities.
    }
    \label{fig:joint_task_effects}
\end{figure*} 
To study our first research question, we examined how network properties influence group performance both overall and for specific task qualities (Fig. \ref{fig:joint_task_effects}).
Like others \cite{Mason2012Collaborative, Meluso2023Multidisciplinary}, we used multivariate ordinary least squares regressions with robust standard errors to isolate the independent performance effects of each network property while controlling for task qualities and agent rationality bounds (see Methods).
Fig. \ref{fig:joint_task_effects}A shows these effects as a heatmap: the first column shows each network property's average performance (the main effects) across all task types.
Subsequent columns show the combined performance of each network property and task quality (task-conditional effects).
For brevity, we refer to ``generate tasks'' to mean tasks with high generate difficulty; similarly for choose, coordinate, and negotiate tasks.
Fig. \ref{fig:joint_task_effects}B illustrates how specific network topologies perform on coordinate and negotiate tasks as functions of network density, decentralization, and path length.
All values represent statistically significant effects ($p<0.05$) in standard deviations above or below the dataset mean.

The main performance effects of network properties are uniformly small with an average magnitude of 0.07 standard deviations.
The main effect of network density is only +0.07 standard deviations, while decentralization (mean eigenvector centrality) shows -0.13 standard deviations, and average shortest path length shows -0.18 standard deviations.
These modest main effects suggest that network properties do not uniformly improve or impair artificial collective performance across all task qualities, similar to findings on human groups \cite{Centola2022Network}.
Rather, their influence depends on the type of task being solved.

This task-dependence becomes evident when examining conditional effects.
The average magnitude of the task conditional effects is 0.33 standard deviations, 4.5 times larger than for the main effects.
For generate tasks---which require finding novel states across complex landscapes---network density provides a substantial benefit of +0.87 standard deviations, while decentralization contributes +0.77 standard deviations.
Shorter average path lengths also help (effect of -0.26 on path length, meaning shorter paths improve performance).

Choose and coordinate tasks exhibit similar network property benefits but with even larger effect magnitudes.
For choose tasks, which involve selecting among multiple alternatives, density provides +0.97 standard deviations benefit and decentralization adds +1.10 standard deviations.
Coordinate tasks, which require synchronizing interdependent actions, show the strongest effects: density contributes +1.60 standard deviations and decentralization contributes +1.84 standard deviations---23 times larger and 14 times larger than their respective main effects.
Shorter path lengths benefit both task types as well, with effects of -0.44 and -0.94 standard deviations respectively.
These patterns indicate that generate, choose, and coordinate task performances improve when interpretive abilities are abundant and widely distributed, as they are for decentralized interpretive networks with efficient communication paths.

Negotiate tasks reveal a contrasting pattern.
In these tasks, agents must balance conflicting assessments of performance.
Unlike other task qualities, negotiate tasks show negative effects for density (-0.55 standard deviations) and decentralization (-0.68 standard deviations), while benefiting from longer average paths (+0.24 standard deviations).
This suggests that negotiate tasks benefit from sparser, more centralized topologies in which a few generalists mediate between specialists.

Fig. \ref{fig:joint_task_effects}B corroborates these patterns through specific network topologies.
For coordinate tasks (left column), dense, decentralized graphs with short path lengths fall further to the right (like the complete, hypercube, and high-probability random graphs).
Most of these graphs show strong positive performance effects (large green circles).
In contrast, negotiate tasks (right column) show a different pattern: sparser networks and those with longer paths---including tree structures and low-probability random graphs---appear in regions corresponding to better performance.
These patterns also replicate across other optimizers with varying effect magnitudes (\textit{SI Appendix S1.1}).

A noteworthy exception to these trends is the complete graph which, despite its high density, decentralization, and short paths, underperforms on coordinate tasks and overperforms on negotiate tasks.
This aligns with convergent findings in collective human intelligence on the inferior performance on efficient networks \cite{Lazer2007Network, Derex2016Partial, Brackbill2020Impact, Centola2022Network}.
However, several mechanisms related to information diffusion speed \cite{Lazer2007Network} and search interference \cite{Meluso2023Multidisciplinary} might cause this phenomenon.
We revisit these mechanisms in the following section when examining how rationality bounds moderate network effects.

Task-specific patterns likely emerge because collective tasks are network games \cite{Galeotti2010Network, Meluso2023Multidisciplinary} that groups ``play'' by collectively searching a state space.
A group's task performance depends on the interdependent actions of individuals which correspond to network topologies.
Tasks aided by state diversity benefit from sparser connectivity that preserves independent search, while tasks emphasizing information aggregation benefit from denser connectivity.
This matches evidence in human groups that network topology has opposite effects on information diffusion and exploration \cite{Lazer2007Network, Shore2015Facts}.
Coordinate tasks illustrate the aggregation case: dense interpretive ties help agents interpret neighbors' actions and make productive adjustments.
Negotiate tasks then illustrate the diversity case: dense interpretive ties may create premature convergence toward suboptimal compromise positions before agents can adequately explore mutually-beneficial alternatives.

\subsection*{Loose bounds favor specialists, tight bounds favor generalists}

\begin{figure*}[b]
    \centering
    \includegraphics[width=7.5in]{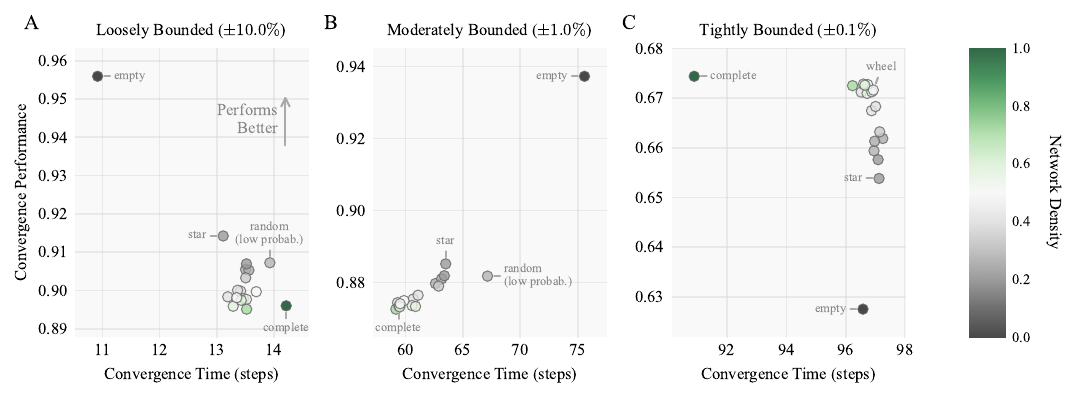}
    \caption{
        \textbf{Network property effects on convergence timing and performance vary across rationality bounds.}
        Each panel shows convergence performance versus convergence time for teams of 8 agents, with points colored by network density (gray = specialists, green = generalists).
        (A) At loose rationality bounds ($\pm10\%$), specialists achieve both faster convergence and superior performance.
        (B) At moderate bounds ($\pm1\%$), a Pareto frontier emerges: specialists converge slower but achieve better performance.
        (C) At tightly bounded rationality ($\pm0.1\%$), this reverses: generalists achieve both fastest convergence and best performance.
        Groups of 2, 4, and 16 agents show similar results.
        Hence, optimal network topologies may depend on the relative capacity of agents to search a state space.
        }
    \label{fig:convergence_performance}
\end{figure*} 
To answer our second research question, we examined how network density affects multi-agent performance across different bounds of rationality (Fig. \ref{fig:convergence_performance}).
These rationality bounds constrain agents' abilities to search the state space by limiting the domain of each decision variable per time step, thereby modeling Simon's concept of bounded rationality \cite{Simon1957Models}.
We refer to these three regimes as loosely bounded ($\pm10\%$), moderately bounded ($\pm1\%$), and tightly bounded ($\pm0.1\%$) search conditions.
We calculated mean convergence performance and convergence time for each network topology and rationality bound.
Examining convergence outcomes across rationality bounds shows that the relationship between network topology, convergence timing, and performance quality differs markedly across rationality bounds with three distinct regimes.

First, when rationality bounds are loose ($\pm10\%$), specialist-focused topologies substantially outperform generalist topologies on both performance and speed (Fig. \ref{fig:convergence_performance}A).
The empty graph---representing pure specialists with no interpretive connections---achieves 6.7\% higher performance than the complete graph (0.956 vs. 0.896) while converging 23\% faster (11 vs. 14 steps).
Between these density extrema, topology averages cluster near the performance of the complete graph, with lower densities slightly outperforming higher densities.

This relationship reverses when rationality is tightly bounded ($\pm0.1\%$; Fig. \ref{fig:convergence_performance}C).
Here, generalist-focused topologies excel: the complete graph outperforms the empty graph by 7.3\% (0.674 vs. 0.628) while converging 6\% faster (91 vs. 97 steps).
At this bound, many topologies approach or reach the 100-step time limit before fully converging, meaning these performance values represent the best states achieved within available computational steps rather than true convergence values.
Still, topologies with density as low as 0.5 (such as wheel graphs) achieve 99.7\% of the complete graph's performance with half the interpretive ties, though with more time required.

The moderately bounded case ($\pm1\%$) exhibits a qualitatively distinct pattern (Fig. \ref{fig:convergence_performance}B): a speed-quality trade-off.
Specialist networks converge 27\% slower than generalist networks (76 vs. 59 steps) but achieve 7\% better performance (0.937 vs. 0.873).
The empty graph stands as an outlier, achieving the highest performance but at substantially greater time cost than all other configurations, which cluster more closely to the complete graph in both timing and performance.
This creates a discontinuous Pareto frontier where design decisions involve a choice between topologies of expensive-but-excellent specialists and faster-but-adequate generalists, rather than a smooth continuum of intermediate options with increasing density.

What drives these outcomes mechanistically?
Rationality bounds determine the local landscape each agent optimizes over during each turn.
At tight bounds ($\pm0.1\%$), agents effectively optimize over smooth, convex regions.
Each additional interpretable neighbor improves an agent's gradient estimation by incorporating more dimensions of the group's objective function.
This provides more accurate directional information for hill-climbing, so generalist topologies (with dense interpretive ties) outperform specialist topologies (with sparse interpretive ties) because agents optimize faster.
With loose bounds ($\pm10\%$), agents often optimize over rugged, non-convex regions containing many local optima.
Each additional interpretable neighbor adds decision variables---and thus dimensions \cite{Bellman1961Adaptive}---to an agent's optimization task, creating exponentially larger decision spaces.
Given the fixed computational budgets imposed by bounded rationality, these high-dimensional spaces become increasingly difficult to sample.
Specialist topologies therefore outperform generalist topologies because agents sample more effectively in lower-dimensional spaces.

These patterns replicate with two other optimizers (L-BFGS-B \& Dual Annealing, \textit{SI Appendix S1.2}).
However, Random Walk agents show weaker effects at tight bounds and reversed patterns at loose bounds (also \textit{SI Appendix S1.2}).
Unlike the other optimizers, which search strategically, Random Walk agents evaluate single random samples.
So, groups of Random Walk agents with more interpretive ties appear to improve their evaluation accuracy without creating the high-dimensional sampling challenges that strategic optimizers face.
This suggests that our mechanistic effects may require goal-directed optimization under bounded rationality instead of random, undirected optimization.

This progression extends both Simon's bounded rationality concept \cite{Simon1957Models} and March's exploration-exploitation framework \cite{March1991Exploration} into multi-agent systems, suggesting that appropriate network topologies can partially compensate for limited computational capacity in complex decision spaces.

\section*{Discussion}

An interpretive network describes which individuals in a group can understand each other's actions.
In these networks, individuals with few interpretive abilities correspond to specialists while those with many function as generalists.
Our results suggest that topological properties of these networks shape the collective performance of multi-agent systems depending on both task qualities and agents' rationality bounds.

Across task qualities, interpretive network properties have small performance effects (average 0.07 standard deviations).
But with respect to specific task qualities, network effects are 4.5 times larger (average 0.33 standard deviations).
Some tasks favor decentralized topologies (generate, choose, coordinate tasks) while others favor centralized ones (negotiate tasks).
Bounded rationality mediates these relationships, though.
When agents have loose rationality bounds relative to a decision space, specialist topologies perform best because they sample state spaces more effectively; with tight bounds, generalists outperform because they optimize faster.
A speed-quality trade-off lies between these extremes.

These findings support a nuanced theory of collective artificial intelligence.
The benefits of human coordination \cite{Riedl2017Teams, Riedl2021Quantifying} rely upon how well social perceptiveness topologically balances information efficiency and information diversity \cite{Lazer2007Network, Woolley2010Evidence, Barkoczi2016Social, Centola2022Network}.
Similarly, our findings show that for most collaborative tasks (generate, choose, and coordinate), agentic systems benefit from interpretive abilities that produce dense, decentralized networks which also balance efficiency and diversity \cite{Meluso2023Multidisciplinary}, though negotiate tasks benefit from centralized topologies with network intermediaries instead.

Our results also yield practical recommendations for crafting efficient multi-agent systems across application domains.
Current approaches to multi-agent system design often select agent abilities and network topologies without systematically considering how they interact with each other, with task qualities, or with agent rationality bounds \cite{Gronauer2022Multiagent, Horling2004Survey}.
Our findings suggest this approach risks substantial performance losses.
Several design principles based on interpretive networks may overcome these challenges.
Like human multidisciplinary teams \cite{Meluso2023Multidisciplinary}, dense topologies where most agents have broad interpretive abilities provide robust performance, particularly across generate, choose, and coordinate tasks.
Centralized configurations with a few coordinating agents become preferable when tasks require reconciling competing objectives (negotiate tasks).
AlphaStar's centralized league topology is consistent with these recommendations \cite{Vinyals2019Grandmaster}.
Generalist main agents trained widely against specialized exploiters, and the system achieved grandmaster-level performance.
However, StarCraft itself involves generate, choose, and negotiate qualities, underscoring the difficulty of designing topologies for single task qualities.
Empirical studies of LLM multi-agent systems also reach compatible conclusions, finding that coordination benefits depend on task structure and degrade when individual agents are already capable \cite{Kim2025Science}.

Regarding agent rationality bounds: when agents can thoroughly search decision spaces (loose rationality bounds), sparse interpretive topologies (such as star graphs) may help agents sample state spaces more thoroughly while preventing agents from interfering with each other's optimization processes.
Conversely, when individual compute is severely constrained relative to the task (tight rationality bounds), fully connected topologies may be unnecessary.
Moderately connected topologies (density $\geq0.5$, like a wheel graph) achieve near-optimal performance with substantially fewer interpretive ties (Fig. \ref{fig:convergence_performance}C).
At moderate rationality bounds, though, a fundamental trade-off emerges between performance and convergence time.
System designers must weigh these competing objectives based on their specific requirements and constraints.

In terms of computational efficiency, our results suggest significant opportunities to optimize resource use and performance through appropriate multi-agent topologies.
Pairing topologies with task qualities could yield large gains.
Decentralized systems show nearly 2 standard deviation improvements on coordinate tasks, while specialist topologies achieve 6.7\% better performance in 23\% less time than generalist topologies at loose rationality bounds.
These improvements hold potential to reduce societal needs for compute time, for more powerful agents, and for greater numbers of agents.
Given the substantial energy and computational costs of multi-agent systems \cite{Economist2024BigTech, IEA2024Electricity, Leppert2025What, IEA2025Energy}, understanding which interpretive network properties provide benefits and costs should be an engineering priority.

Furthermore, these findings hold implications for human-AI collaboration, a major focus of current research \cite{Riedl2025AI}.
To date, the many complications of human-computer interaction mean mixed human-AI groups often underperform the better of the two working alone \cite{Vaccaro2024When}.
The interpretive network topologies that benefit artificial collectives may inform optimal configurations for hybrid teams as well, potentially bridging the performance characteristics of human and artificial collective intelligence.
Interpretive networks could provide a common language for designing mixed topologies of humans and artificial agents, positioning each according to their respective interpretive strengths.

Finally, our methodology enables future research on coordination processes \cite{Riedl2021Quantifying}, dynamic network adaptation \cite{Galesic2023Collective}, and heterogeneous group composition \cite{Page2019Diversity} in artificial collectives.
By bridging human and artificial collective intelligence research, this work advances our understanding of how interpretive network properties shape emergent intelligence across organizational scales and contexts.

Several limitations warrant future investigation.
Our model employed static interpretive network topologies, while adaptive networks that evolve based on performance or task qualities often yield additional benefits \cite{Rubenstein2014Programmable, Galesic2023Collective}.
We also explored only homogeneous agent optimization abilities, whereas heterogeneous groups with differentiated specializations and rationality bounds are likely to provide further insights and better performance \cite{Hong2004Groups, Page2019Diversity}.
While our findings point to interpretive network properties as key determinants of performance, factors like initial condition sensitivity and system size likely contribute to these patterns, meriting further investigation.
Although partially validated by large language model findings \cite{Kim2025Science}, validation with other types of agents and in specific domains like scientific discovery, systems engineering, and autonomous systems represents important next steps toward practical application.

\appendix
\small

\section*{Methods}
\label{sec:methods}
We designed a systematic experiment with multi-agent systems to investigate how agent interpretive abilities, rationality bounds, and task qualities influence collective performance.
To this end, we carefully designed properties of our multi-agent systems, simulation experiments, and statistical analyses (Fig. \ref{fig:methods}).
Code accompanying this paper can be found at \url{https://github.com/meluso/mas-interpretive-networks-code/}. Data can be found on Zenodo at \url{https://doi.org/10.5281/zenodo.19682737}.

\subsection*{Multi-Agent Systems}

For our multi-agent systems, we made design decisions at the group level and the agent level.

\subsubsection*{Groups}

At the group level, we represented collective configurations of interpretive abilities through network topologies and objective functions.
Network topologies ranged from completely disconnected agents with no interpretive abilities (empty graphs) to fully connected agents with the ability to interpret all other actions (complete graphs), plus 16 intermediate topologies including small world, random, and preferential attachment networks (\textit{SI Appendix S2.1}).
Including many topologies enabled us to significantly vary values of network properties with known effects elsewhere and therefore to measure their relative performance effects here (see \textit{SI Appendix S2.2} for network properties and corresponding metrics).
We also varied group sizes ($n=2$, 4, 8, 16 agents) to control for effects of group size on network properties and individual degrees of freedom while holding system computational resources constant.
This was done by assigning each agent a set of variables to optimize that is inverse in size to the group size ($v=16$, 8, 4, or 2 variables, respectively), totaling 32 variables per system in all cases.

Multi-agent systems were assigned task objective functions $h$ that define collective performance mappings.
Each agent $i$ operates with a neighborhood objective function $g_i$ that matches $h$'s mathematical form but incorporates only information the agent can interpret through its network connections: its own assigned variables plus those of its connected neighbors.
For example, if the group task $h$ averages all agents' variables, then each agent evaluates $g_i$ as the average of only its own variables and its neighbors' variables.
This design ensures that interpretive network topology directly determines which information agents use to evaluate states.

\subsubsection*{Agents}

At the agent level, we abstractly modeled artificial agents as optimizers that iteratively search constrained state spaces.
We use classical optimization algorithms as scientific representations of artificial agents---specifically the Nelder-Mead simplex algorithm (shown in the main Results) plus the L-BFGS-B algorithm, a simulated annealing algorithm, and a random walk algorithm for comparison (\textit{SI Appendix S1}) \cite{Virtanen2020SciPy}.

We chose this approach because it captures a fundamental property shared across many multi-agent systems while providing scientific tractability.
Artificial agents generally must search for high-quality states within computational and design constraints---whether through gradient descent, beam search, policy networks, or tree traversal, they are bounded by finite resources and processing capacities \cite{Simon1957Models, Russell1994Provably, Gottwald2019Systems}.
Classical optimizers model this functional constraint directly: agents evaluate candidate states and select among them within defined limits.
This abstraction does not require modeling the specific implementation details of particular AI architectures, enabling us to isolate the effects of interpretive network properties on collective performance in a scientifically rigorous way.

In our multi-agent algorithm, agents iteratively optimize their assigned variables at each time step, incorporating information from network neighbors before producing new states (see Algorithm).
This design enables us to study how network properties mediate collective search under bounded rationality while maintaining relevance to contemporary and future agentic systems that operate under computational bounds.

We operationalized bounded rationality through four rationality bounds ($b=\pm0.1\%$, $\pm1\%$, $\pm10\%$, $\pm100\%$) representing the fraction of each decision variable's domain that an agent can search per time step.
These bounds span a range of realistic constraint levels:
Bounds of $\pm0.1\%$ model tightly constrained agents that search within a small fraction of the decision space.
Bounds of $\pm1\%$ and $\pm10\%$ model moderately and loosely constrained agents, respectively.
Bounds of $\pm100\%$ model unbounded agents, unconstrained in their ability to search the full decision space.
This parameterization facilitates systematic investigation of how collective performance depends on both network topology and the degree of individual rationality bounds.

The mechanisms we identify---how interpretive abilities shape information access and influences collective search behavior---apply broadly to systems where agents evaluate state quality within constraints, though specific quantitative effects may vary with implementation details.

\subsubsection*{Algorithm}

During each simulation trial, agents iteratively optimize their assigned variables within their rationality bounds.
The procedure follows this sequence:

\begin{enumerate}
    \item Each agent $i \in I$ observes the current state (variable values $x_j$) of connected neighbors $j \in J_i$ as determined by the interpretive network topology.
    \item Agent $i$ uses its optimizer to generate a candidate state $x'_i$ within its rationality bounds, based on its objective function $g_i(x_i, \{x_j\}_{j \in J_i})$.
    \item After all agents generate candidate states simultaneously, agents calculate a new objective evaluation $g'_i$ using the updated states $\{x'_j\}_{j \in J_i}$ of all connected neighbors.
    \item The group objective $h$ is computed using all agents' new states $(x'_1, \ldots, x'_n)$, so $h(\{x'_i\}_{i \in I})$.
    \item The process repeats until convergence (change in $h < 10^{-4}$ for 3 consecutive steps) or maximum iterations (100 steps).
\end{enumerate}

This synchronous update procedure ensures that all agents optimize simultaneously based on the information they can interpret as defined.
The formal algorithmic procedure is detailed in \textit{SI Appendix S4}.

\subsection*{Simulation Experiment}

To systematically examine the effects of different properties, we ran simulation experiments with multi-agent systems, varying the group size, network type, and rationality bounds as described above.
Multi-agent systems were assigned 30 mathematically defined optimization tasks that vary in difficulty along four task quality dimensions from collective intelligence research \cite{McGrath1984Groups, Woolley2010Evidence, Riedl2021Quantifying, Meluso2023Multidisciplinary}:

\begin{itemize}
    \item \textbf{Generate}: Finding novel high-quality states
    \item \textbf{Choose}: Selecting among alternatives
    \item \textbf{Coordinate}: Synchronizing interdependent actions
    \item \textbf{Negotiate}: Balancing competing objectives
\end{itemize}

Tasks range from simple functions like averaging to complex optimization landscapes like the Ackley function (shown in Fig. \ref{fig:methods}).
All objective functions are normalized to $[0, 1]$ to enable comparison across tasks, with 1 representing optimal performance.
Detailed task definitions and difficulty measurement methods are provided in \textit{SI Appendix S3}.

The experimental design crossed 4 group sizes $\times$ 18 network topologies $\times$ 30 tasks $\times$ 4 rationality bounds, with 250 independent trials per parameter combination, yielding about 1.7 million simulation trials.

\subsection*{Statistical Analysis}

We estimated the effects of network properties and topologies on group performance through ordinary least squares regressions with robust standard errors (type HC2).
To isolate the independent contributions of network properties from the confounding effects of multicollinearity (see \textit{SI Appendix S5} for correlation analysis), we conducted separate regression analyses for individual network properties (each with a corresponding network metric) and for specific network topologies.

We used data that includes all 13 network metrics, which required excluding disconnected topologies where path-based metrics (shortest path length, diameter) are undefined.
This filtering retained topologies with valid values for all network properties, providing the most comprehensive characterization of network property effects.
Prior to regression, all continuous variables were normalized using MinMax scaling to the range [0,1].

Each regression included controls for agent rationality bounds (log$_{10}$(agent\_steplim)) and the four task qualities (generate, choose, negotiate, coordinate).
We conducted two variants of each regression: one with main effects only, and one that included interaction terms between the primary predictor and each task quality (task-conditional effects).
This approach allowed us to estimate both the overall effect of each network property or network topology and how that effect varies across different task types.

For network property analyses, we ran separate regressions with each of 13 network metrics as the primary predictor: degree centrality (mean and standard deviation), betweenness centrality (mean and standard deviation), eigenvector centrality (mean and standard deviation), nearest neighbor degree (mean and standard deviation), clustering coefficient, density, degree assortativity, shortest path length (mean), and graph diameter.
These analyses used data from all group sizes (2, 4, 8, and 16 agents) with normalized group size included as a control variable.
Model fitting used scikit-learn's LinearRegression \cite{Pedregosa2011Scikitlearn}, with robust standard errors calculated using the HC2 estimator \cite{Seabold2010Statsmodels}.

For topology-specific analyses, we ran separate regressions for each of the 18 network topologies, using dummy variables to indicate topology with the complete graph as the reference category.

The normalized variables enable interpretation of regression coefficients as standardized effects: a coefficient of 1.0 indicates that a one standard deviation increase in the predictor corresponds to a one standard deviation change in convergence performance.
The intercept represents expected performance when all predictors are at their mean values.

For visualization purposes, we filtered results of Fig. \ref{fig:joint_task_effects}A to include only statistically significant effects ($p < 0.05$).
Task-conditional effects shown in the figure represent the cumulative effect of a network property on a specific task quality, calculated as the sum of the network property's main effect, the task quality's main effect, and their interaction effect, all from the model including interactions.
This cumulative effect captures the total predicted impact of the network property when that specific task quality is at its maximum value while other task qualities are at their mean values.
Analyses for Fig. \ref{fig:joint_task_effects}B used only groups of size 8.
We did this to help visualize the relationships between performance and three network properties (density, decentralization, and average shortest path length) while holding these network metrics constant for each topology.

\bibliographystyle{pnas-new}
\bibliography{library}

\end{document}